\documentclass[apjl]{emulateapj}

\usepackage{natbib}
\usepackage{amsmath}
\usepackage{amssymb}
\usepackage[bf, sf, FIGTOPCAP, nooneline, tight]{subfigure}
\usepackage{graphicx}
\usepackage{dcolumn}
\usepackage{bm}
\usepackage[usenames,dvipsnames]{color}
\usepackage[colorlinks=true, linkcolor=BrickRed, citecolor=Blue, urlcolor=Blue, filecolor=Blue]{hyperref}
\usepackage{booktabs}
\usepackage{hhline}

\def\beq{\begin{equation}}
\def\eeq{\end{equation}}
\def\ber{\begin{eqnarray}}
\def\eer{\end{eqnarray}}

\def\om{\Omega_{0m}}
\def\atridot{\stackrel{...}{a}}
\def \lleq {\lower0.9ex\hbox{ $\buildrel < \over \sim$} ~}
\def \ggeq {\lower0.9ex\hbox{ $\buildrel > \over \sim$} ~}
\def\apj{{Astroph.\@ J.\ }}
\def\mn{{Mon.\@ Not.\@ Roy.\@ Ast.\@ Soc.\ }}

\def\prd{{Phys.\@ Rev.\@ D\ }}

\def \jetpl {JETP Lett.\ }
\def\etal{{\it et al.}}

\def\apj{ApJ}%
\def\apjl{ApJ}%
\def\apjs{ApJS}%
\def\ao{Appl.~Opt.}%
\def\aap{A\&A}%
\def\prd{Phys.~Rev.~D}%

\def\deg{\ifmmode^\circ\else$^\circ$\fi}

\begin{document}

\title{Model independent evidence for Dark Energy Evolution from\\ Baryon Acoustic Oscillations}

\author{V. Sahni\altaffilmark{1}, A. Shafieloo\altaffilmark{2,3}, A. A. Starobinsky\altaffilmark{4,5}}

\altaffiltext{1}{Inter-University Centre for Astronomy and Astrophysics, Post Bag 4, Ganeshkhind, Pune 411~007, India}
\altaffiltext{2}{Asia Pacific Center for Theoretical Physics, Pohang, Gyeongbuk 790-784, Korea}
\altaffiltext{3}{Department of Physics, POSTECH, Pohang, Gyeongbuk 790-784, Korea}
\altaffiltext{4}{Landau Institute for Theoretical Physics RAS, Moscow 119334, Russian Federation}
\altaffiltext{5}{Kazan Federal University, Kazan 420008, Republic of Tatarstan, Russian Federation}

\email{varun@iucaa.ernet.in}
\email{arman@apctp.org}
\email{alstar@landau.ac.ru}

\begin{abstract}
Baryon Acoustic Oscillations (BAO) allow us to determine the expansion history of the Universe, thereby shedding 
light on the nature of dark energy. Recent observations of BAO's in the SDSS DR9 and DR11 have provided us with 
statistically independent measurements of $H(z)$ at redshifts of 0.57 and 2.34, respectively. We show that these 
measurements can be used to test the cosmological constant hypothesis in a model independent manner by means of 
an improved version of the $Om$ diagnostic. Our results indicate that the SDSS DR11 measurement of 
$H(z) = 222 \pm 7$ km/sec/Mpc at $z = 2.34$, when taken in tandem with measurements of $H(z)$ at lower redshifts, 
imply considerable tension with the standard $\Lambda$CDM model. Our estimation of the new diagnostic $Omh^2$ 
from SDSS DR9 and DR11 data, namely $Omh^2 \approx 0.122 \pm 0.01$, which is equivalent to 
$\Omega_{0m}h^2$ for the spatially flat $\Lambda$CDM model, is in tension with the value 
$\Omega_{0m}h^2 = 0.1426 \pm 0.0025$ determined for $\Lambda$CDM from Planck+WP.
This tension is alleviated in models in which the cosmological constant was dynamically
{\em screened} (compensated) in the past. 
%An example of such a model is provided and it is shown that the screening mechanism can 
Such evolving dark energy models display a pole in the effective equation of state
of dark energy at high redshifts, which emerges as a {\em smoking gun} test for these
theories. 
\end{abstract}


\section{Introduction}

There is ample observational evidence to suggest that the expansion of the universe
is accelerating, fuelled perhaps by {\em dark energy} (DE)
 which violates the strong 
energy condition, so that $\rho + 3P < 0$. While the cosmological constant with
$8\pi G T_{ik} = \Lambda g_{ik}$ and $P = -\rho \equiv -\Lambda/8\pi G$, 
envisioned by Einstein almost
a century ago, fulfills this requirement, the tiny value associated with $\Lambda$
has prompted theorists to look for alternatives in which dark energy evolves
with time including modified gravity \citep{Sahni_Starobinsky2000,Carroll2001,Peebles_Ratra2003,Paddy2003,Sahni2004,Copeland_2006,ss06,clifton,Shafieloo2014}. %(Cosmological acceleration may also be sourced by theories
%which depart from GR at late times.)

Meanwhile, the very simplicity of the cosmological constant has prompted the
search for {\em null-diagnostics} which
can inform us, on the basis of observations, whether or not DE is the cosmological constant.

One such diagnostic is the {\em Statefinder} $r = \atridot/aH^3$ (also called
the jerk) whose value stays pegged to unity {\em only in} $\Lambda$CDM \citep{statefinder1,statefinder2} (also see\citep{chiba98,V04}).
Thus if observations were to inform us that $r \neq 1$, then this would 
imply a falsification of the cosmological constant hypothesis.

A second null diagnostic, $Om(z)$, is defined as \citep{om,zunkel_clarkson}
\beq
Om(z) = \frac{\tilde{h}^2(z) - 1}{(1+z)^3 - 1}~, ~~~~\tilde{h} = H(z)/H_0~.
%Om(x_2,x_1) = \frac{\tilde{h}^2(x_2) - \tilde{h}^2(x_1)}{x_2^3-x_1^3}, ~~~~ x = 1+z~,
\label{eq:om}
\eeq
%where $\tilde{h} = H(z)/H_0$. 
A remarkable feature of $Om$ is that its value
remains pegged to $\om$ in $\Lambda$CDM. In all other DE models the value of $Om(z)$
{\em evolves with time}.

While the Statefinder has proved exceedingly versatile in differentiating between
rival DE models, a distinguishing feature of $Om$ is that it depends only
upon the expansion rate, $H(z)$, and is therefore easier to determine
from observations than $r$ (also see \citep{om3,V04,chiba00,maryam}).
$Om$ can also be written as a two-point diagnostic \citep{om3}
\beq
%Om(z_2;z_1) = \frac{\tilde{h}^2(z_2)-\tilde{h}^2(z_1)}{(1+z_2)^3 - (1+z_1)^3}, %~~h(z) = H(z)/H_0
Om(z_i;z_j) = \frac{\tilde{h}^2(z_i)-\tilde{h}^2(z_j)}{(1+z_i)^3 - (1+z_j)^3},
\label{eq:om1}
\eeq
with $Om(z;0)$ defined in (\ref{eq:om}).
Consequently, if the Hubble parameter is known at two or more redshifts then 
$Om(z_i;z_j)$ can be reconstructed and one can address the issue of whether
DE is the cosmological constant or not. Recent observations of BAO's in the SDSS catalogue have paved the way for 
reconstructing $Om$ by determining statistically independent values of $H(z)$ at several redshifts \citep{bao_2014}. Using their determination of 
$H(z=2.34) = 222 \pm 7$ km/sec/Mpc, \citet{bao_2014} reported a surprising
 2-2.5$\sigma$ tension with the predictions of standard $\Lambda$CDM with best-fit
 Planck  
parameters. In this paper we revisit this 
inconsistency using a null diagnostic approach involving an improved version of
$Om$.  We affirm the results of \citet{bao_2014} and also demonstrate that 
{\em screened} models of dark energy provide a better fit to the BAO data
than $\Lambda$CDM. (It may be appropriate to mention that \citet{bao_2014} is
a preprint and it
is possible that these results be revised prior to publication.)

\section{Data, Method \& Results}

An advantage of using BAO's to deduce the nature of DE is that the former are
measured on large scales and hence determined primarily by the linear regime
of gravitational instability, a theory that has been meticulously developed and 
studied over the past
several decades.
In this letter we reconstruct $Om$ using recent determinations
of $H(z)$ and attempt to answer the question as to whether DE behaves like the
cosmological constant.
Consider first, the following small improvement of $Om$ which yields large dividents.
Multiplying both sides of (\ref{eq:om1}) by $h^2$ where
 $h = H_0/100$km/sec/Mpc, results in the
{\em improved $Om$ diagnostic}
\beq
Omh^2(z_i;z_j) = \frac{h^2(z_i) - h^2(z_j)}{(1+z_i)^3 - (1+z_j)^3}, %~~~~h(z) = 
%H(z)/100 km/sec/Mpc~.
\eeq
where $h(z) = H(z)/100$km/sec/Mpc.
A significant advantage of $Omh^2$ is that, for $\Lambda$CDM:
\beq
Omh^2 = \om h^2~.
\label{eq:concordance}
\eeq
Since observations of the CMB inform us that \citep{planck} $\om h^2 = 0.1426 \pm 0.0025$,
 it follows that for the cosmological constant $\Lambda$:
\beq
Omh^2 = 0.1426 \pm 0.0025~.
\eeq
Consequently, a departure of $Omh^2$ from the above value would signal that DE {\em is not}
 $\Lambda$.
As we shall show, this is precisely what is suggested by the recent measurement
of $H(z) = 222 \pm 7$ km/sec/Mpc at $z = 2.34$ made on the basis of BAO's in the $Ly\alpha$
forest of BOSS DR11 quasars \citep{bao_2014}.

One notes that for $n$ independent measurements of $H(z_i)$, $z_i \in z_1 \cdots z_n$,
the pairwise diagnostic $Omh^2(z_i;z_j)$ can be determined in $\frac{n(n-1)}{2}$ different ways.
In the present case $n=3$, which leads to 3 independent measurements of $Omh^2(z_i;z_j)$, namely
\ber
Omh^2(z_1;z_2) &=& 0.124 \pm 0.045\nonumber\\
Omh^2(z_1;z_3) &=& 0.122 \pm 0.010\nonumber\\
Omh^2(z_2;z_3) &=& 0.122 \pm 0.012\nonumber\\
\label{eq:omh2}
\eer
where $z_1 = 0, z_2 = 0.57, z_3 = 2.34$, and the Hubble parameter at these redshifts
is $H(z=0) = 70.6 \pm 3.3$ km/sec/Mpc \citep{efstathiou_2014}, $H(z=0.57) = 92.4 \pm 4.5$ km/sec/Mpc \citep{bao_2013} and
$H(z=2.34) = 222 \pm 7$ km/sec/Mpc \citep{bao_2014}.

One notes from (\ref{eq:omh2}) that the {\em model independent} value of $Omh^2 \simeq 0.12$ is quite stable,
and is in tension with the $\Lambda$CDM-based value
$Omh^2\vert_{\Lambda{\rm CDM}} \simeq 0.14$.
For the pair $Omh^2(z_1;z_3)$ and $Omh^2(z_2;z_3)$ the tension with $\Lambda$
 is at over $2 \sigma$.

We should note here that these results are quite robust and not unduly sensitive to the value of $H(z=0)$. Assuming $H(z=0)=73.8 \pm 2.4$ km/sec/Mpc,
 which is the best estimated value by~\citet{riess2011}, results in $Omh^2(0;2.34)=0.121 \pm 0.009$. While using $H(z=0)=67.1 \pm 1.2$,
 which is the best fit value for Hubble parameter from Planck concordance $\Lambda$CDM model,
 results in $Omh^2(0;2.34)=0.123 \pm 0.009$. Hence it is clear that the `final' value
 of $H(z=0)$ should not affect the derived value of $Omh^2$ significantly, which
 suggests that our results for this quantity are robust. Likewise using the more recent SDSS galaxy BAO DR10 and DR11 result of $H(z=0.57) = 96.8 \pm 3.4$ km/sec/Mpc \citep{bao_2013B} we get $Omh^2(z_2;z_3) = 0.120 \pm 0.010$ which is in agreement
 with our earlier estimations of $Omh^2$. This is mainly due to
the high precision measurement of $H(z=2.34)$ which makes the determination of
$Omh^2$ less sensitive to the value of $H(z)$ at lower redshifts. 

%One might note that a null diagnostic such as $Omh^2$, relying as it does only on the expansion
%history, $H(z)$, has a significant advantage over the equation of state, $w(z)$, which
%has been frequently used to study cosmic acceleration. Since
%\beq
%w(z) = \frac{(2x/3)d \log H/dx - 1}{1-(H_0/H)^2\om x^3}~, ~~~~ x = 1+z
%\eeq
Thus far our treatment has been model independent and we have refrained from commenting
on the physical implications of the SDSS measurements of $H(z)$. However,
as already noted in \citet{bao_2014}, these implications can be quite serious.
Indeed the expansion rate at $z=2.34$,
namely $H(z=2.34) = 222 \pm 7$ km/sec/Mpc \citep{bao_2014}, could be in tension
not only with $\Lambda$CDM but with DE models based on the general relativistic equation
($\kappa = 8\pi G/3$)
\beq
H^2(z) = \kappa\lbrack \rho_{\rm DE}(z) + \rho_{0m} (1+z)^3\rbrack~~~
{\rm with} ~\rho_{\rm DE}(z) \geq 0~.
\label{eq:GR}
\eeq
%Substituting $H(z=2.34) = 222 \pm 7$ km/sec/Mpc and 
%$H_0 = 70.6 \pm 3.2$ km/sec/Mpc into (\ref{eq:GR}),
%one finds $H^2(z)/H^2_0 < \om (1+z)^3$ at $z=2.34$ unless $\om < 0.265$.
%
%This might imply one of the following:
%(i) $\om < 0.265$, (ii) $\om > 0.265$ but $\rho_{\rm DE} < 0$ at high $z$,
%(iii) the framework (\ref{eq:GR}) is inadequate since one might be dealing here
%with a {\em modified gravity} theory.

Note that by setting $\rho_{\rm DE} = 0$ in (\ref{eq:GR})
one finds
\beq
\frac{h^2(z)}{(1+z)^3} = \om h^2.% ~~~~h(z) =
%\frac{H(z)}{100 km/sec/Mpc}~.
\eeq
Substituting $H(z=2.34) = 222 \pm 7$ km/sec/Mpc one obtains
$h^2(z)/(1+z)^3= 0.132 \pm 0.008$ which is somewhat {\em lower} than the CMB based value
$\om h^2= 0.142 \pm 0.002$.

This might imply one of the following: (i) $\rho_{\rm DE}(z) < 0$ at high $z$ \citep{bao_2014,cardenas},
(ii) there is non-conservation of matter so that (\ref{eq:GR}) does not hold,
(iii) the framework (\ref{eq:GR}) is inadequate since one could be dealing 
with a {\em modified gravity} theory.

An example of (iii) is provided by models in which dark energy, and in particular
the cosmological constant, is {\em screened} (or 
compensated) by a dynamically evolving counter-term.
In the case of the latter, eqn. (\ref{eq:GR}) is modified to
%of \citet{ss02} in which
\beq
H^2(z) = \frac{\Lambda}{3} + \kappa\rho_{0m} (1+z)^3
- f(z)~, ~~~~ f(z) > 0~.
\label{eq:brane}
\eeq
Examples of this behaviour may be found in: (i) theories in which $\Lambda$ relaxes from
a large initial value via an {\em adjustment mechanism} \citep{dolgov,bauer},
(ii) in cosmological models based on Gauss-Bonnet gravity \citep{copeland}, and
(iii) in Braneworld models \citep{ss02}, etc. More generally, this behaviour occurs in modified
gravity (e.g. in scalar-tensor gravity) when the effective gravitational constant 
$G_{eff}(z)<G_{eff}(0)\equiv G$ , if we {\em define} $\rho_{DE}(z)$ using the present value of 
$\kappa$ in Eq. (\ref{eq:GR}) following \cite{beps00,ss06}.

A key feature of such models is that if $f(z)$ grows monotonically with redshift
(but at a slower rate than $(1+z)^3$ in order to preserve the matter-dominated regime),
then a stage will come when $\Lambda/3$ is exactly balanced by 
$f(z)$, resulting in $H^2(z_*) \simeq \kappa\rho_{0m} (1+z_*)^3$.
At $z_*$ the {\em effective} equation of state of dark energy, $w(z)$, develops a 
pole, at which $|w(z_*)| \to \infty$. This is easily seen from the expression \citep{ss06}
\beq\label{eq:state}
w(x) = {2 q(x) - 1 \over 3 \left( 1 - \Omega_{\rm m}(x) \right)}
\equiv \frac{(2 x /3) \ d \ {\rm ln}H \ / \ dx - 1}{1 \ - \ (H_0/H)^2
\Omega_{m0} \ x^3}\,\,,
\eeq
where $x=1+z$, $\Omega_{\rm m}(x)= \Omega_{0m}x^3H_0^2/H^2(x)$ and
$q$ is the deceleration parameter.
One finds from (\ref{eq:brane}) and (\ref{eq:state}) that
at $f(z_*) = \Lambda/3$
\beq
w(z_*) = - \frac{(1+z_*)f'(z_*)}{H^2(z_*) - \kappa\rho_{0m}(1+z_*)^3}~.
\eeq
 In other words, $w(z_*)$ diverges when $f(z_*) = \Lambda/3$ and
$H^2(z_*) \simeq \kappa\rho_{0m} (1+z_*)^3$, provided $f'(z_*) \neq 0$.
%that $w(z)$ diverges when $\Omega_{\rm m}(z) \simeq 1$.

As a specific example of a model with this behaviour,
consider the Braneworld model proposed in \citet{ss02} 
and described, in a spatially flat universe, by the equations:
\ber\label{eq:hubble_brane}
{H^2(z) \over H_0^2} &=& \Omega_\Lambda + \om (1\!+\!z)^3 \nonumber\\
&&+ 2 \Omega_l - 2 \sqrt{\Omega_l}\, \sqrt{\om (1\!+\!z )^3 + \Omega_\Lambda + \Omega_l }~, \nonumber\\
\Omega_{\Lambda} &=& 1-\om+2\sqrt{\Omega_l}\,\,,
\eer
where the densities $\Omega$ are defined as :
\beq \label{eq:omegas}
\om = {\rho_{0m} \over 3 m^2 H_0^2} , ~ \Omega_\Lambda = {\Lambda \over 3 m^2H_0^2} ,
~ \Omega_l = {1 \over l_c^2 H_0^2} .
\eeq
$l_c = m^2/M^3$ is a new length scale ($m$ and $M$ refer
respectively to the four and five dimensional Planck masses),
and $\Lambda$ is the brane tension associated with a 3 dimensional brane embedded
in a 4+1 dimensional bulk space-time.

\begin{figure*}[!t]
\centering
\begin{center}
\vspace{-0.00in}
\centerline{\mbox{\hspace{0.in} \hspace{1.4in}  \hspace{1.4in} }}
$\begin{array}{@{\hspace{0.0in}}c@{\hspace{0.5in}}c@{\hspace{0.5in}}c}
\multicolumn{1}{l}{\mbox{}} &
\multicolumn{1}{l}{\mbox{}} \\ [0.5cm]
\includegraphics[scale=0.45, angle=-90]{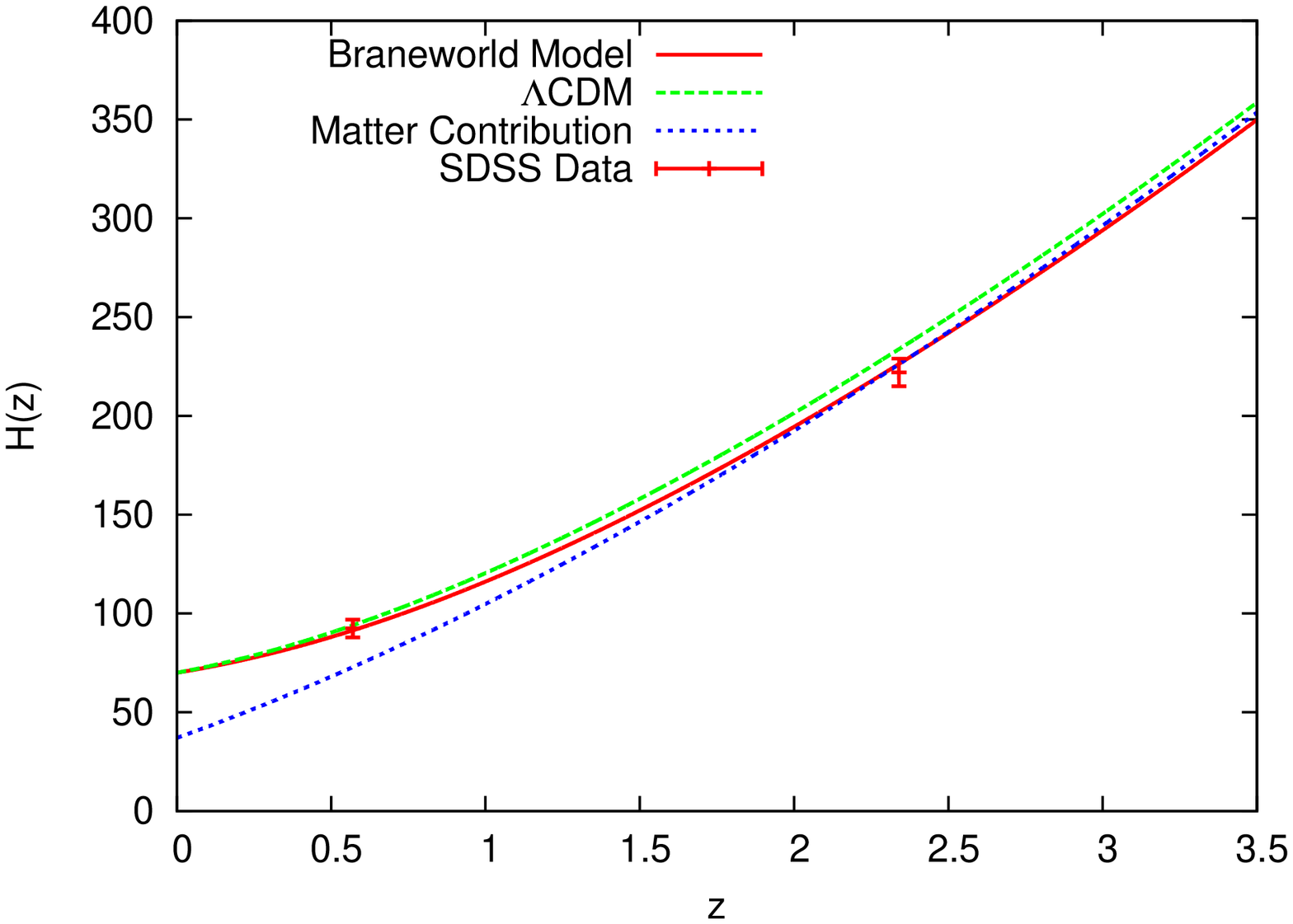}
\includegraphics[scale=0.45, angle=-90]{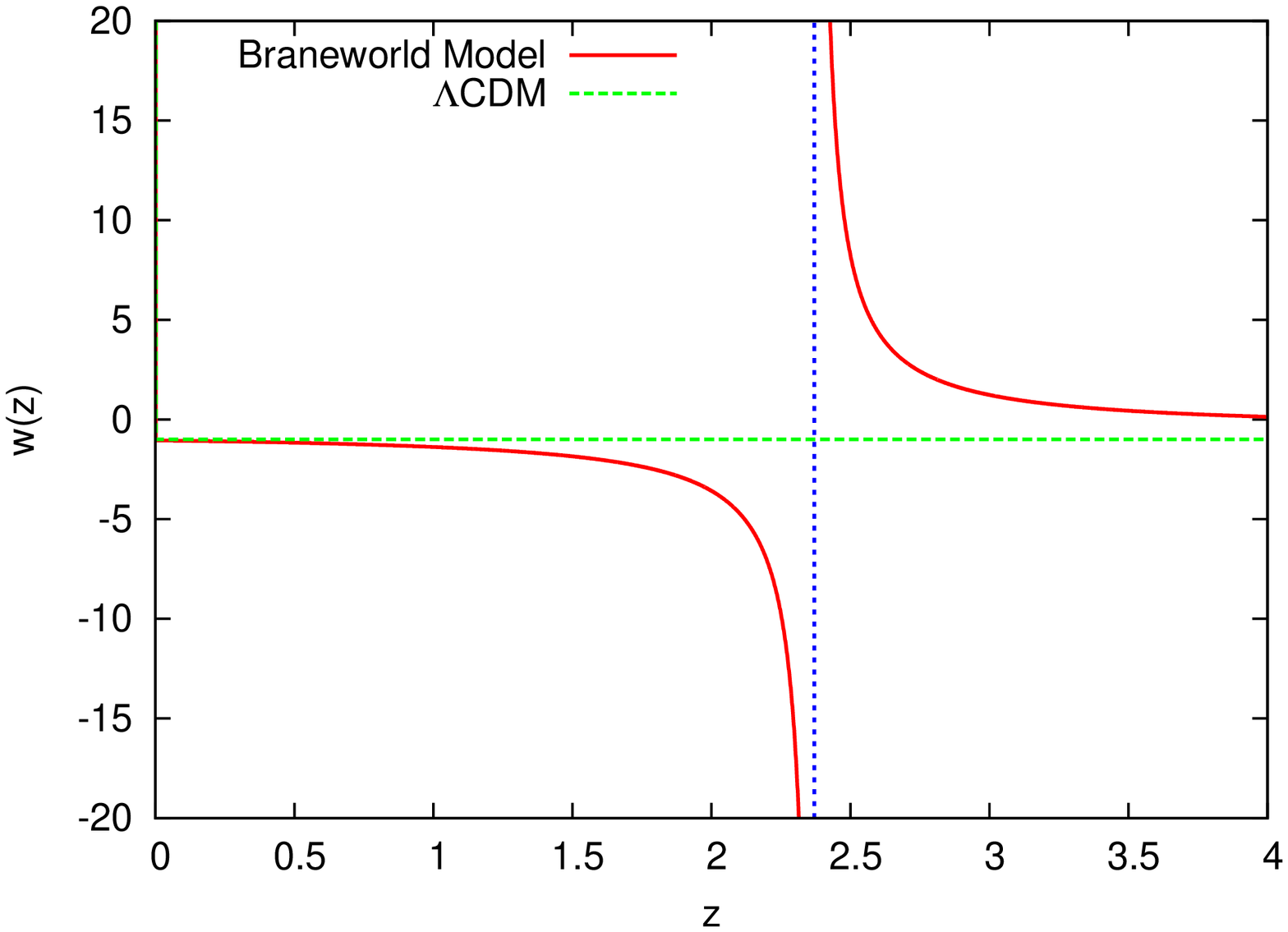}
\end{array}$
\end{center}
\caption{The Hubble parameter (left panel) and the effective 
equation of state of dark energy (right panel) are shown 
for the Braneworld model described by (\ref{eq:hubble_brane})
 (solid red) and $\Lambda$CDM (dotted green).
Also shown is the matter contribution: $H_0\sqrt{\om (1+z)^3}$
where $H_0 = 70$ km/sec/Mpc and $\om = 0.28$ (dotted blue).
In the Braneworld model the cosmological constant is {\em screened} in the past as
a result of which
the expansion rate
 drops {\em below} that in $\Lambda$CDM at high $z$. This feature permits the Braneworld to better
account for the low value of $H(z=2.34)$ discovered in \citet{bao_2014}. 
Note that $H_{\rm Brane} \simeq H_0\sqrt{\om (1+z)^3}$ at $z \simeq 2.4$.
The associated pole in $w(z)$ at $z \simeq 2.4$ is shown in the right panel.
The parameters for the Braneworld model are $\om = 0.28$ and $\Omega_\ell = 0.025$
in (\ref{eq:hubble_brane}).}
\label{fig1}
\end{figure*}

As shown in figure \ref{fig1} the expansion rate in this model
can drop below that in $\Lambda$CDM at high $z$. It can therefore better account for the
lower-than-anticipated value for $H(z=2.34)$ discussed in \citet{bao_2014}. Note also
the pole in $w(z)$ at $z \simeq 2.4$. It might be mentioned that the presence of the pole
in this model does not signal any pathologies since $w(z)$
is an {\em effective} equation of state. This is also true for the other theoretical
models in which $w(z)$ exhibits a pole \citep{bauer,copeland}.
Note that a pole in the equation of state may be possible to pick out in
future 
type Ia supernova (SNIa) data sets using model independent reconstruction, as demonstrated
in \citet{arman2006}.
Finally one might point out that although
dark energy in the Braneworld behaves like a phantom it does not share 
the latter's pathologies \citep{ss02,sahni2005}. The model also
 agrees with SNIa observations \citet{alam}.

A detailed analysis of models with screened/compensated dark energy will be the
subject of a future work.%\altaffilmark{6}\altaffiltext{6}
%{One might note in passing that the spatially
%flat `coasting' model with $a(t) \propto t$ suggested in \cite{melia07}
% as an alternative to $\Lambda$CDM
%is in serious tension with $H(z)$ observations.
%This can be seen from the fact that $H(z) = H_0 (1+z)$ in such a model, which,
%for $H_0 = 70$km/sec/Mpc, results in $H(z = 0.57) = 110$ km/sec/Mpc. The latter is much 
%larger than the observed value of $H(z = 0.57) = 92.5 \pm 4.5$ km/sec/Mpc.}

There is another important issue that requires elaboration. The derived value of $H(z=2.34)$ given 
by~\citet{bao_2014} is scaled at $r_d=147.4$ Mpc from the Planck+WP fitting of concordance 
cosmology, where $r_d$ is the sound horizon at the drag epoch. One may argue that playing with the 
parameter, $r_d$, may help reconcile the concordance model with data. However this cannot 
 be true since the value of $r_d$ used to derive $H(z=2.34)$ has been obtained
 assuming $\Lambda$CDM and the discrepancy between $Omh^2$ and $\om h^2$
obtained by us is also based on $\Lambda$CDM cosmology -- see (\ref{eq:concordance}). One should however note that it is possible to lower the value of $r_d$ by increasing the expansion rate in the early Universe through the inclusion of
 an extra relativistic species. But this would imply a departure from the minimal standard $\Lambda$CDM model (though not in its dark energy sector). 

Its also important to point out that a lower (than in $\Lambda$CDM) value of $H(z)$ at
high $z$ would affect the growth of matter density perturbations, perhaps speeding
them up relative to $\Lambda$CDM. Indeed, on scales much smaller than the horizon
and within the framework of general relativity, linearized perturbations
are described by the equation \citep{peebles80}
\beq
{\ddot \delta} + 2H{\dot\delta} - 4\pi G\bar{\rho}\delta = 0~.
\label{eq:pert}
\eeq
Clearly a lower value of $H(z)$ results in a suppression of the damping term 
$2H{\dot\delta}$ (relative to $\Lambda$CDM) and therefore to a faster growth in 
$\delta$. This could have important implications for structure formation
which will soon be probed to great depth and accuracy by SKA, LSST, etc.
However, (\ref{eq:pert}) generically does not hold in modified gravity theories.
In particular, in scalar-tensor gravity this equation has formally the same form at sufficiently 
small scales but with the effective gravitational constant $G_{eff}(t)$ instead of $G$ 
\citep{beps00}. Therefore a detailed analysis of perturbation growth in such models needs to be 
carried out before firm predictions can be made about $\delta(z)$.

%For this combination (fitting Planck+WP with concordance model) we get $\Omega_{0m}h^2 = 0.1426 \pm 0.0025$ which is very much different from our estimated value of $Omh^2$ using $H(z=2.34$ data. This inconsistency cannot be resolved by changing the value of $r_d$. 

%We should also note here that 
 
%Smaller values of $H_0 \simeq 67.3$ permit a somewhat larger value $\om < 0.088$,
%but the problem remains.
%\beq
%\frac{H^2(z)}{H^2_0} = \frac{\rho_{\rm DE}}{\rho_{\rm cr}^{(0)}} + \om (1+z)^3~,
%\eeq
%where $\rho_{\rm cr}^{(0)} = 3H_0^2/8\pi G$,

\section{Summary}
To summarise, this short letter demonstrates that the recent estimation of $H(z=2.34)$ from BAO observations in the SDSS DR11 data is in tension with CMB observations assuming standard $\Lambda$CDM. This tension is independent of the current value of 
the Hubble parameter $H(z=0)$. In our analysis we have implemented an improved version of the
 $Om$ 
diagnostic, called $Omh^2$, which can be derived by having independent
 measurements of $H(z)$ at two redshifts.
$Omh^2$ should be equal to $\Omega_{0m} h^2$ if the universe corresponds to
 spatially flat $\Lambda$CDM. Our estimated value of $Omh^2 \approx 0.122 \pm 0.01$ 
(which should also be the value of $\Omega_{0m} h^2$ for $\Lambda$CDM) is robust against 
variations of the Hubble parameter $H_0$ and is in strong tension with $\Omega_{0m}h^2 = 0.1426 \pm 0.0025$ given by Planck+WP.

In the absence of systematics in the CMB \& SDSS data sets, our results suggest a 
strong tension between concordance cosmology and observational data. 
Since resolving this discrepancy by changing initial conditions and/or the
 form of the primordial spectrum might be difficult (note that $\Omega_{0m} h^2$ 
does not change much if one deviates smoothly from the power-law form of the primordial spectrum~\citep{Hazra:2013nca,Hazra_Shafieloo2014,Hazra:2014aea}), 
allowing dark energy to evolve seems to be the most plausible approach to this problem.  
Evolving dark energy models which might accommodate the SDSS data better than $\Lambda$CDM include those in which
the cosmological constant was {\em screened} in the past.
The effective equation of state in such models develops a pole at high $z$,
 which emerges as
 a {\em smoking gun} test for such scenarios.

\begin{acknowledgments}
A.S. wishes to acknowledge support from the Korea Ministry of Education, Science and Technology, Gyeongsangbuk-Do and Pohang City for Independent Junior Research Groups at the Asia Pacific Center for Theoretical Physics. A.S. would like to acknowledge the support of the National Research Foundation of Korea (NRF-2013R1A1A2013795). A.A.S. was partially supported by the grant RFBR 14-02-00894 and by the Scientific Programme ``Astronomy'' of the Russian Academy of  
Sciences.
\end{acknowledgments}

\end{document}